\documentclass[11pt]{article}
\usepackage{graphicx}
\usepackage{feynmf}
\usepackage{epsfig}
\usepackage{here}
\usepackage{wrapfig}
\usepackage{psfrag}
\usepackage{lscape}
\usepackage{rotating}

\newcommand{\BABARPubYear}    {02}

\newcommand{\BABARConfNumber} {029}
\newcommand{\SLACPubNumber} {9314}

\newcommand{\verysmallrule}{\rule[-2.0mm]{0.0cm}{0.8cm}}
\newcommand{\vvsmallrule}{\rule[-1.0mm]{0.0cm}{0.6cm}}
 
\input pubboard/babarsym

\long\def\inst#1{\par\nobreak\kern 4pt\nobreak
    {\it #1}\par\vskip 10pt plus 3pt minus 3pt}

\begin{document}
{\pagestyle{empty}

\begin{flushright}
\babar-CONF-\BABARPubYear/\BABARConfNumber \\
SLAC-PUB-\SLACPubNumber \\
July 2002 \\
\end{flushright}

\begin{center}
\end{center}

\par\vskip 2cm

\begin{center}
\Large \bf Measurement of the First Hadronic Spectral Moment from Semileptonic $B$ Decays 
\end{center}
\bigskip

\begin{center}
\large The \babar\ Collaboration\\
\mbox{ }\\
July 25, 2002
\end{center}
\bigskip \bigskip

\begin{center}
\large \bf Abstract
\end{center}
A preliminary determination of the first moment of the hadronic mass distribution $\langle M_X^2 - \overline{m}_D^2 \rangle$  in semileptonic $B$ decays
has been obtained as a function of the minimum lepton momentum, ranging from 0.9 to 1.6 \gevc. The measurement is based on a new technique involving \BB\ events in which one fully reconstructed $B$ meson decays hadronically and the recoiling $B$  decays semileptonically. 
The mass of the hadrons in the semileptonic decay is determined from a kinematic fit to the whole event.  For different minimum lepton momenta, the mass distribution is decomposed into contributions from various charm resonant states and a non-resonant contribution, allowing for the determination of the first moment.
From these moments the Heavy Quark Effective Theory (HQET) parameters $\lambda_1$ and $\bar{\Lambda}$ can be derived. 
For lepton momenta in the \B\ rest frame above 1.5\gevc, we find a first moment that is compatible with existing measurements. However, if we extend the measurement to lower values of lepton momenta, the data can only be described by Operator Product Expansion calculations if we use significantly different values for $\bar\Lambda$ and $\lambda_1$ than obtained from earlier measurements based on lepton momentum spectra and the photon spectrum in $b\to s\gamma$ transitions.

\vfill
\begin{center}
Contributed to the 31$^{st}$ International Conference on High Energy Physics,\\ 
7/24---7/31/2002, Amsterdam, The Netherlands
\end{center}

\vspace{1.0cm}
\begin{center}
{\em Stanford Linear Accelerator Center, Stanford University, 
Stanford, CA 94309} \\ \vspace{0.1cm}\hrule\vspace{0.1cm}
Work supported in part by Department of Energy contract DE-AC03-76SF00515.
\end{center}

\newpage
} 

\begin{center}
\small

The \babar\ Collaboration,
\bigskip

B.~Aubert,
D.~Boutigny,
J.-M.~Gaillard,
A.~Hicheur,
Y.~Karyotakis,
J.~P.~Lees,
P.~Robbe,
V.~Tisserand,
A.~Zghiche
\inst{Laboratoire de Physique des Particules, F-74941 Annecy-le-Vieux, France }
A.~Palano,
A.~Pompili
\inst{Universit\`a di Bari, Dipartimento di Fisica and INFN, I-70126 Bari, Italy }
% ...

\end{center}\newpage

\section{Introduction}
\label{sec:Introduction}

The heavy quark limit in QCD has become a very useful tool for relating inclusive $B$ decay properties, like the semileptonic branching fraction, 
to the charged current couplings, $|V_{cb}|$ and  $|V_{ub}|$. 
These inclusive observables  can be calculated using expansions in powers of the strong coupling constant $\alpha_s(m_b)$ and in inverse powers of the $B$ meson mass, $m_B$, that include non-perturbative parameters $\bar{\Lambda}$, $\lambda_1$ and $\lambda_2$. From the $B^*-B$ mass splitting $\lambda_2$ has been determined to be $0.128 \pm 0.010 \gev^2$ and it is expected that $\bar{\Lambda}$ and $\lambda_1$ will be calculated with techniques such as lattice QCD. However, they can also be extracted experimentally from inclusive measurements.  

In this paper we report a measurement of the invariant mass $M_{X}$ of the hadronic system recoiling against the charged lepton and the neutrino in semileptonic $B$ decays. In the framework of 
HQET, calculations of the  moment $\langle M_X^2 - \overline{m}_D^2 \rangle$ have been carried out \cite{Falk:1998jq} using an Operator Product Expansion (OPE) in inverse powers of the $B$ meson mass up to order $\alpha^2_s \beta_0$ and $1/m_B^3$. Here $\overline{m}_D = (m_D + 3 m_{D^*})/4 = 1.975~\gevcc$ is the spin-averaged $D$ meson mass and $\beta_0=(33-2n_f)/3=25/3$ is the leading order QCD $\beta$ function.
In these expansions the non-perturbative parameters $\bar{\Lambda}$ (${\cal O}(1/m_B)$), $\lambda_1$ and $\lambda_2$ (${\cal O}(1/m^2_B)$) appear. A measurement over a large region of phase space can test the validity of the assumptions made in the OPE, such as parton-hadron duality.
 
Our determination of the hadronic moment in semileptonic \B\ decays exploits new capabilities made possible by very large samples of reconstructed \B\ mesons, in contrast to earlier measurements by CLEO~\cite{Cronin-Hennessy:2001fk}. The data sample for the analysis relies on \BB\ events in which one \B\ meson is fully reconstructed and thus the momentum of the semileptonically decaying $B$ is known. This allows us to assign the remaining tracks and clusters in the event to the hadronic mass $M_X$ of the semileptonic decay. Furthermore, the determination of $M_X$ can be substantially improved by application of a two-constraint kinematic fit to the full event.

The moment $\langle M_X^2 - \overline{m}_D^2 \rangle$ of the $M_X$ distribution is measured for lepton momenta (calculated in the $B$ rest-frame) as low as 0.9 \gevc. The wide momentum range improves the sensitivity of the measurement to a variety of 
hadronic final states, including higher resonances of the $D$ and non-resonant $D^{(*)}\pi$ combinations.
\\

\section{The \babar\ Detector and Data Set}
\label{sec:babar}

Our analysis is based on data recorded with the \babar\ detector \cite{Aubert:2001tu} at the \pep2\ energy-asymmetric \epem storage ring at SLAC. 
The detector consists of a five-layer silicon vertex tracker (SVT), a 40-layer drift chamber
(DCH), a detector of internally reflected Cherenkov light (DIRC), and an
electromagnetic calorimeter (EMC) assembled from 6580 CsI(Tl) crystals, all embedded in a solenoidal magnetic field of 1.5 T and surrounded by an instrumented flux return (IFR).
The data sample corresponds to an integrated luminosity 
of 51\invfb that was collected at the \FourS\ resonance.

\section{Analysis Method}
\label{sec:Analysis}

The basic analysis technique is based on \BB\ events in which one of the $B$ mesons decays hadronically and is fully reconstructed ($B_{reco}$) and the recoiling $B$ ($B_{recoil}$) decays semileptonically. 
The requirement of such a decay topology allows the \B\ momentum to be determined, so that transformations to the rest frame of the recoiling \B\ meson are possible. It also results in a clean sample of \BB\ events, with the flavor of the reconstructed \B\ meson determined.

\subsection{Selection of Hadronic $B$ Decays, $B \ra D Y$}

The reconstruction of $B$ decays is designed 
to identify samples of decays of the type $B \rightarrow D Y$. Here $D$ refers to a charm meson and $Y$ represents a collection of hadrons of total charge $\pm 1$, composed of $\pi^{\pm}, K^{\pm}, \KS$ and $\piz$ mesons.  The charm meson serves as a ``seed'' for the selection of candidates. We use four different seeds: $D^+$ and $D^{*+}$ for $\Bzb$ and $D^0$ and $D^{*0}$ for $B^-$ (charge conjugate modes are implied throughout this document). Several different decay modes are used, each of which is characterized by a signal-to-background ratio dependent on the multiplicity and on the composition of the $Y$ system. In events with more than one reconstructed $B$ decay, the decay mode with the highest {\it a priori} signal-to-background ratio is selected. 

$B_{reco}$ candidates are identified by two kinematic variables,
the beam energy-substituted mass, $M_{ES}=\sqrt{E^{*2}_{beam} - p_B^{*2}}$,
and the energy difference in the \FourS\ rest frame, 
$\Delta E = E_B^* - E^*_{beam}$. Here $E^{*}_{beam}$ is the beam energy,
and $p_B^{*}$ and $E_B^*$ refer to the $B$ momentum and energy in the center-of-mass frame. $B_{reco}$ candidates are required to have $\Delta E$ within three standard deviations of zero and $M_{ES} > 5.27 \gevcc$. A sideband region in $M_{ES}$ used for background subtraction is defined by $5.20 < M_{ES} < 5.27 \gevcc$. 
An example for the $M_{ES}$ distribution can be found in Figure~\ref{fig:mes}.

\subsection{Selection of Semileptonic Decays, $B \ra X \ell \nu$}

Semileptonic $B$ decays are identified by the presence of one  and only one electron or muon above a minimum momentum $P^*_{min}$ measured in the rest frame of the $B$ meson recoiling against the $B_{reco}$. $P^*_{min}$ is 
varied for the different measurements in the range of $0.9 -1.6 \gevc$.
This requirement on the lepton  momentum reduces backgrounds from secondary charm or $\tau^{\pm}$ decays, and from hadrons faking leptons.

To further improve the background rejection, we require that the lepton charge and the flavor of the reconstructed $B^{\pm}$ be consistent with a prompt semileptonic decay of the $B_{recoil}$.  For a reconstructed \Bz we do not impose this restriction, so as not to lose events due to $\Bz - \Bzb$ mixing.

Electrons are identified using a likelihood-based algorithm, combining the track momentum with the energy, position, and shape of the shower measured in the EMC, the Cherenkov angle and the number of photons measured in the DIRC, and the specific energy loss in the DCH.
The efficiency of these selection requirements has been measured with radiative Bhabha events as a function of the laboratory momentum $p_{lab}$ and polar angle $\theta_{lab}$, and has been corrected for the higher multiplicity of \BB events using Monte Carlo simulations. The average electron selection efficiency is ($90 \pm 2$)\%.

Muons are identified using a cut-based selection for minimum ionizing tracks, relying on information from the finely segmented instrumented flux return of the magnet. The number of interaction lengths traversed by the track, the spatial width of the observed signals, and the match between the IFR hits and the extrapolated charged track are used in the selection. 
The muon identification efficiency has been measured with $\mu^+\mu^-
(\gamma)$ events and two-photon production of $\mu^+\mu^-$ pairs. 
The average muon selection efficiency is $(70\pm4)\%$.

The misidentification probabilities for pions,  kaons, and protons  have been extracted from  selected control samples in data. They vary between 
0.05\% and 0.1\% (1.0\% and 3\%) for the electron (muon) selection.

To further reduce backgrounds we require the total charge of the event to be $|Q_{tot}|= |Q_{Breco} + Q_{Brecoil}| \leq 1$, and restrict the total missing mass to $|M_{miss}^2| < 1.0~[\gevcc]^2$.
We explicitly allow for a charge imbalance to reduce the dependence on the exact modeling of charged particle tracking in the Monte Carlo simulation, especially at low momenta, and the production of tracks from photon conversions.

\begin{figure}[t]
\begin{center}
 \mbox{\epsfig{file=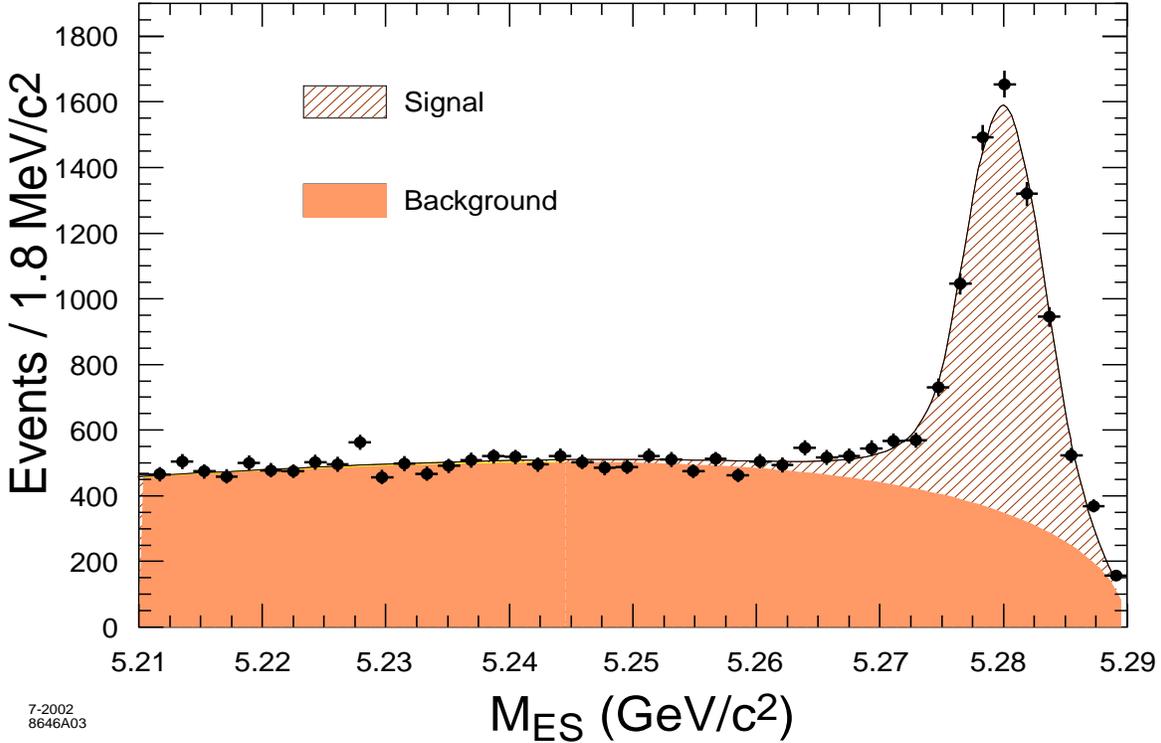,height=10cm,width=15.5cm}}
\caption{\em The $M_{ES}$ distribution for 
selected events with one reconstructed $B$ decay and a lepton with $P^*_{min}=0.9 \gevc$.
\label{fig:mes} }
\end{center}
\end{figure}

Figure ~\ref{fig:mes} shows the $M_{ES}$ distribution for events with a $B_{reco}$ candidate and a charged lepton candidate from the recoiling $B$ meson. A sideband subtraction is performed in bins of $M_X$ to remove the background contribution under the peak at the $B$ mass. For this purpose a sum of background (approximated by a function first introduced by the ARGUS Collaboration~\cite{Albrecht:1987nr}) and signal (described by a function first used by the Crystal Ball experiment~\cite{Skwarnicki:1986xj}) is fitted to the $M_{ES}$ distribution of the data. For $M_{ES} > 5.27 \gevcc$ we find $5819$ signal events and $3585$ background events. The fit to the $M_{ES}$ distribution is shown in Figure ~\ref{fig:mes}. 
\subsection{Reconstruction of the Hadronic Mass $M_X$}
  
\label{subsec:kfit}

The hadron system in the decay $B \ra X \ell \nu$ is formed from $K^{\pm}$, $\pi^{\pm}$, and photons that are not associated with the $B_{reco}$ candidate or identified as leptons. Specifically, we select charged tracks in the fiducial volume 
$0.41 < \theta_{lab} < 2.54~{\rm rad}$ with a minimum transverse momentum $p_t > 120 \mevc$. For photons we require an energy $E_{lab}>100 \mev$ and $0.41 < \theta_{lab} < 2.409~{\rm rad}$.

To improve the resolution on the measurement of the hadronic mass $M_X$, we exploit the kinematic constraints of the $\BB$ state by performing a kinematic fit to the full event using the measured momenta and energies of all particles.
The measured four-momentum $P_X^m$ of the $X$ system can be written as

\begin{equation}
P_X^m = \sum_{i=1}^{N_{ch}} P^{ch}_i  
      + \sum_{j=1}^{N_{\gamma}} P^{\gamma}_j 
\end{equation}
\noindent
where $P$ are four-momenta and the indices $ch$ and $\gamma$ 
refer to the selected charged tracks and photons, respectively. Depending on particle identification the charged tracks are assigned either the $K^{\pm}$ or $\pi^{\pm}$ mass.

The missing momentum in the event is reconstructed by relating the sum of the four-momenta of the colliding beams, $Q_{CM}$, 
to the measured four-momenta of the $B_{reco}$ candidate, the $X$ hadrons, and the charged lepton,
\begin{equation}
P_{miss}=Q_{CM}-P_{reco}^m-P_X^m - P_{\ell}^m .
\end{equation}
The measured invariant mass squared, $M_{miss}^2 = P_{miss}^2$, is an important discriminant on the quality of  the reconstruction of the total recoil system. 
Any secondary particles from the decay of the hadronic $X$ system that are undetected or poorly measured will impact the measurement of both $M_X$ and $M_{miss}^2$.  Likewise any sizable energy loss of the leptons via bremsstrahlung or internal radiation will impact the measurement of these two quantities.  The effect of initial state radiation is rather small, due to the narrow width of the \FourS\ resonance. 
\begin{figure}[t]
\begin{center}
\mbox{\epsfig{file=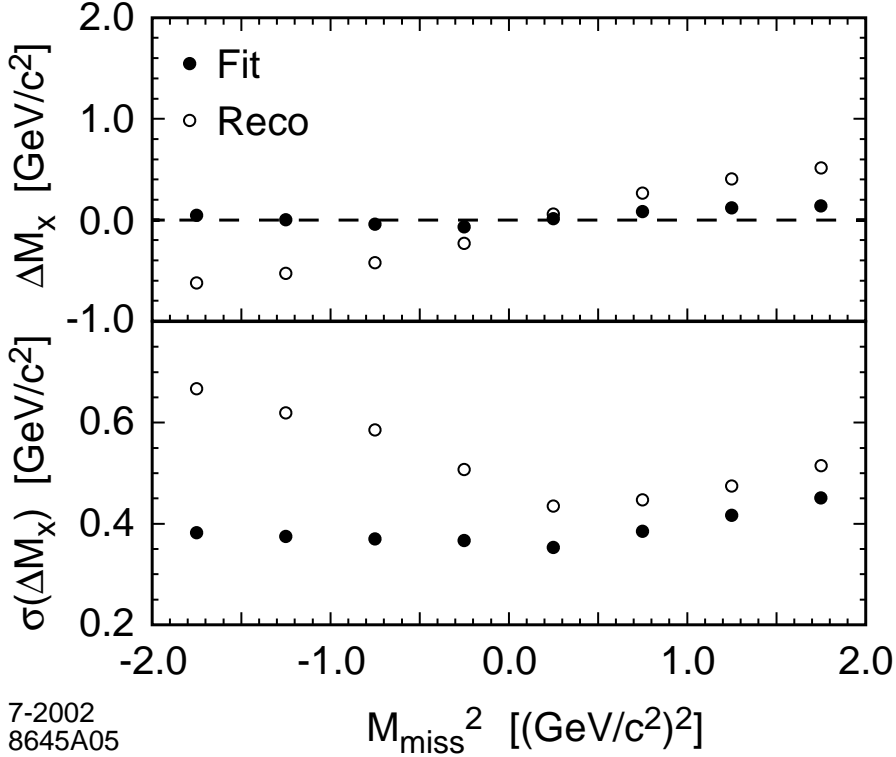,scale=1.5}}
\caption{\em Monte Carlo comparison of the reconstructed and fitted mass $M_X$:  Mean and r.m.s. of the distribution $\Delta M_X= M_X^{True}-M_X^{Reco}$ (open circles) and $\Delta M_X= M_X^{True}-M_X^{Fit}$ (dots) as a function of the measured missing mass squared $M_{miss}^2$.
\label{scanMmiss_vcb} }
\end{center}
\vspace{-0.7cm}
\end{figure}
We exploit the available kinematic information from the full event, namely the $B_{reco}$ and the $B_{recoil}$ candidate, by performing a 2C kinematic fit that imposes four-momentum conservation, the equality of the masses of the two $B$ mesons, $M_{recoil}=M_{reco}$, and forces $M_{miss}^2 = M_{\nu}^2 = 0$.
The fit takes into account event-by-event the measurement errors of all individual particles and the measured missing mass. 
\unitlength1.0cm 
\begin{figure}[t]
\begin{center}
\begin{picture}(15.,10.)
\put(-1.2,0.0){\mbox{\epsfig{file=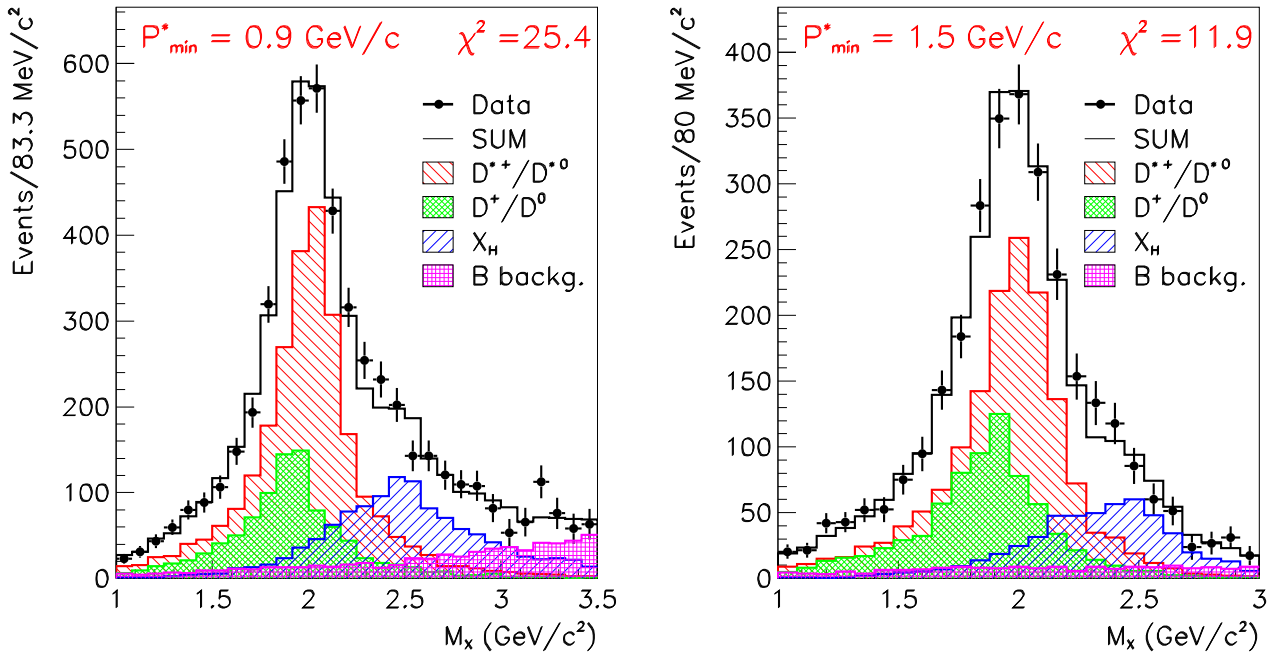, scale=.8}}}
\end{picture}
\end{center}
\vspace{-0.5cm}
\caption{\em Sideband-subtracted $M_X$ distribution for $P^*_{min}=0.9 \gevc$ (left) and $P^*_{min}=1.5\gevc$ (right). The hatched histograms show the contributions from $B \rightarrow D^* \ell \nu$, $B \rightarrow D \ell \nu$ and $B \rightarrow X_H \ell \nu$ decays as determined by the fit, as well as the background distribution. The white histogram represents the sum of all the aforementioned distributions. 
The histograms extend over the range in  $M_X$ that was used in the fit to extract $\langle M_X^2 - \overline{m}_D^2 \rangle$.}
\label{fig:mxdist}
\end{figure}

Figure \ref{scanMmiss_vcb} shows a comparison for $B\ra X_c \ell \nu$ events between the reconstructed and the kinematically fitted mass $M_X$ as a function of the measured $M_{miss}^2$, in terms of the mean and r.m.s. of the distributions for $\Delta M_X= M_X(true)-M_X$.

The constraints of energy and momentum conservation and the
equal-mass hypothesis lead not only to a significant improvement 
in the mass resolution of the $X$ system but also provide an almost unbiased estimator of the mean and a resolution that is largely independent of $M_{miss}^2$.\\

\section{Determination of the Hadronic Mass Moments}

Figure \ref{fig:mxdist} shows the measured $M_X$ distribution of the selected events after sideband subtraction,
for two different minimum lepton momenta $P^*_{min}$. 
Several $B \rightarrow X_c l\nu$ decays 
contribute to this distribution. The dominant decays are $B \rightarrow D^* l\nu$ and  $B \rightarrow D l\nu$, but there are also contributions from decays to higher mass charm states, $D^{**}$ resonances with a mass distribution $X_H^{reso}$ peaked near 2.4 \gevcc, and non-resonant $D^{(*)} \pi$ final states with a broad mass distribution $X_H^{nreso}$ extending to the kinematic limit.

\begin{table}[t]
\begin{center}
\caption[]{\label{tab:mom} {\it Preliminary results for $\langle M_X^2-\overline{m}_D^2 \rangle$ for different values of $P^*_{min}$. The errors are statistical and total systematic respectively. The last four columns show separately the four dominant contributions to the systematic uncertainty: $X_H$ model dependence, detector resolution, background contribution, and sideband (SB) background subtraction.\\}}
\vspace{-0.5cm}
\begin{tabular}{|c||c c c||c|c|c|c|}
\hline
\verysmallrule $P^*_{min}$ ($\gevc$) & \multicolumn{7}{|c|}{$\langle M_X^2-\overline{m}_D^2 \rangle~([\gevcc]^2)$} \\\hline
& & stat. & sys. &$X_H$ model& Detector& Background &SB Subtraction\\\hline\hline
\vvsmallrule 0.9 & 0.694 &$\pm$ 0.081 &$\pm$ 0.080 &0.059 &0.031  &0.030& 0.033\\
\vvsmallrule 1.0 & 0.620 &$\pm$ 0.078 &$\pm$ 0.069 &0.051 &0.026  &0.027& 0.027\\
\vvsmallrule 1.1 & 0.590 &$\pm$ 0.075 &$\pm$ 0.058 &0.043 &0.022  &0.023& 0.024\\
\vvsmallrule 1.2 & 0.555 &$\pm$ 0.073 &$\pm$ 0.049 &0.031 &0.021  &0.020& 0.024\\
\vvsmallrule 1.3 & 0.481 &$\pm$ 0.070 &$\pm$ 0.035 &0.020 &0.021  &0.016& 0.011\\
\vvsmallrule 1.4 & 0.403 &$\pm$ 0.071 &$\pm$ 0.031 &0.014 &0.022  &0.014& 0.009\\
\vvsmallrule 1.5 & 0.354 &$\pm$ 0.074 &$\pm$ 0.030 &0.009 &0.024  &0.011& 0.009\\
\vvsmallrule 1.6 & 0.234 &$\pm$ 0.069 &$\pm$ 0.029 &0.003 &0.026  &0.008& 0.009\\ \hline\hline
\end{tabular}

\end{center}
\end{table}
The observed $M_X$ distribution is fitted to the sum of four distributions 
that we derive from Monte Carlo simulations:  three contributions for  the decays $B \rightarrow D^* l\nu$, $B \rightarrow D l\nu$ and $B \rightarrow X_H l\nu$, and  one for backgrounds.  The ratio of the total branching fractions for the production of $X_H^{reso}$ and $X_H^{nreso}$ is fixed to 1.25 as assumed in the Monte Carlo event generator. The contribution of the different components to the $M_X$ distribution varies as a function of the charged lepton momentum.  The shape and contribution of the background also vary as a function of $P^*_{min}$ and are fixed to the prediction from the Monte Carlo simulation.
The dominant background is due to secondary semileptonic charm decays. Contributions from lepton (primarily muon) misidentification are much smaller. The backgrounds decrease significantly for higher $P^*_{min}$.

Detailed Monte Carlo simulations are used to reproduce the impact of the efficiencies on signal and background and to derive the shape of the signal and background contributions to the $M_X$ distribution. These simulations take into account variations of the detector performance and the beam background.
The decays  $B \rightarrow D \ell\nu$ and $B \rightarrow D^{**} \ell\nu$ are simulated using the ISGW2 model~\cite{ISGW2}, whereas for the decays $B \rightarrow D^* l\nu$ a parameterization of HQET-derived form factors~\cite{Duboscq:1996mv} is used. The non-resonant final states such as $D\pi$ and $D^* \pi$ are modeled according to the prescription of Goity and Roberts \cite{J.GoityandW.Roberts}. 

A binned $\chi^2$-fit to the $M_X$ distribution is performed to determine the relative normalization of the three signal contributions, $f_{D^*}$, $f_D$, and $f_{X_H}$. 
Figure \ref{fig:mxdist} shows the fit results for a minimum lepton momentum of $P^*_{min} = 0.9 \gevc$ and $P^*_{min} = 1.5 \gevc$.
At higher lepton momenta, the contribution from the higher mass states is reduced, specifically the
non-resonant charm states contribute very little above $1.5 \gevc$. 
For $P^*_{min}= 0.9 \gevc$ decays to $D^*$, $D$, and higher mass final states $X_H^{reso}$ and $X_H^{nreso}$ contribute approximately 50\%, 20\%, 20\% and 10\% of the total $M_X$ distribution. 
These fractions change to 60\%, 20\%, 15\% and 5\% for $P^*_{min}=1.5 \gevc$. 
Thus the average $\langle M_X \rangle$ value decreases for higher lepton momenta.

\begin{figure}[H]
\begin{center}
\begin{picture}(15.,15.5)
\put(0.0,0.0){\mbox{\epsfig{file=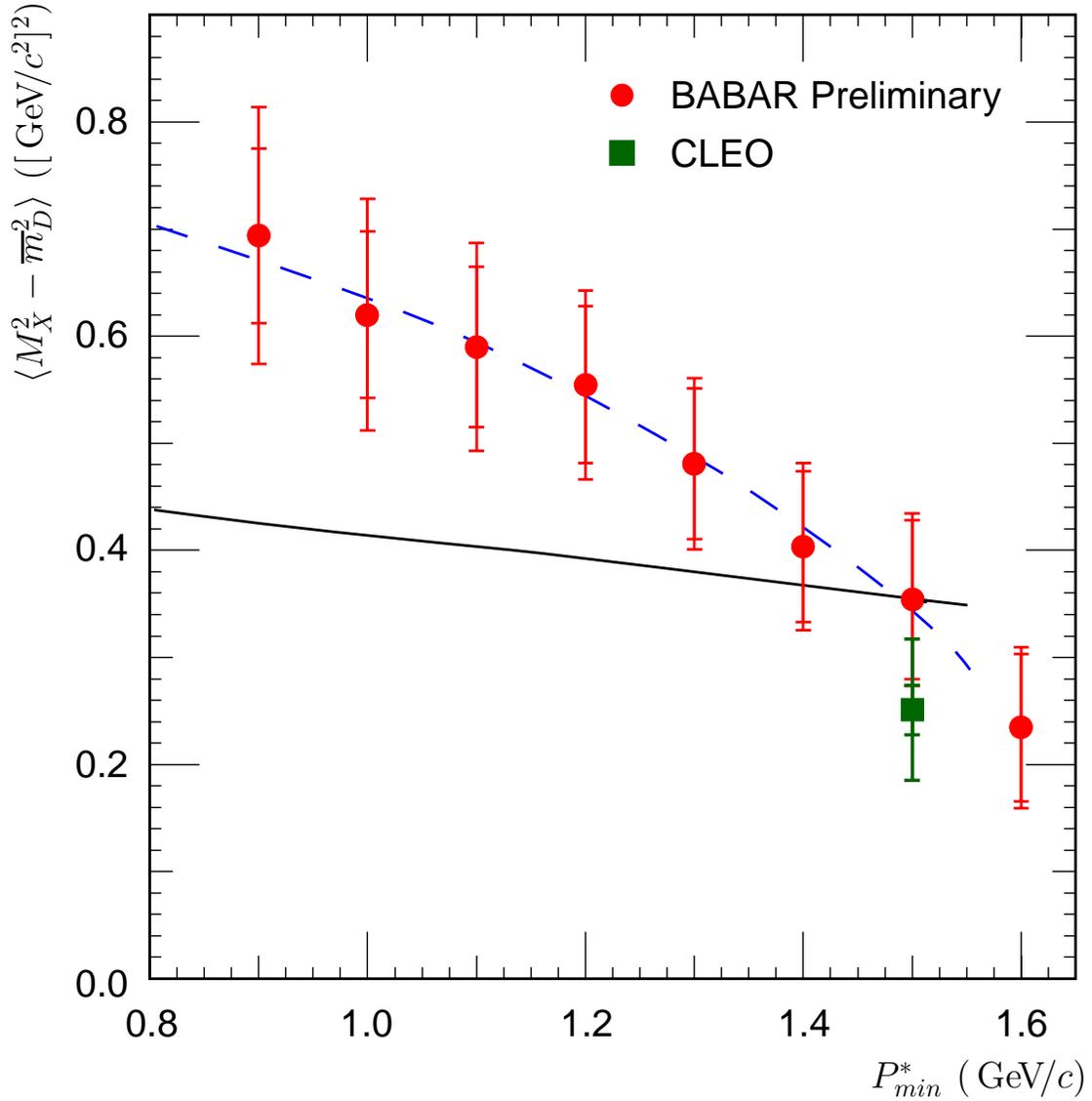, scale=1.05}}}
\put(0.3,10.0){\mbox{\begin{turn}{90} {\bf \Large$\langle M_X^2-\overline{m}_D^2 \rangle~([\gevcc]^2)$}\end{turn}}}
\put(12.0,0.4){{\bf \Large $P^*_{min}~(\gevc)$}}
\end{picture}   
\end{center}
\vspace{-0.7cm}
\caption{\em \label{moments_ivar2} Measured values for $ \langle M_X^2-\overline{m}_D^2 \rangle$ for different minimum lepton momenta, $P^*_{min}$. The inner error bars indicate the statistical uncertainty, the total error bars show the quadratic sum of the statistical and systematic errors. The errors of the individual measurements  are highly correlated. The CLEO measurement~\cite{Cronin-Hennessy:2001fk} is shown for comparison. The lower line shows the predicted variation of the moments for the parameters $\lambda_1 = -0.17~\gev^2$ and $\bar{\Lambda} = 0.35~\gev$ \cite{Chen:2001fj}, that were obtained from the single measurement at $P^*_{min} = 1.5~\gevc$. 
For comparison, the dashed curve shows an example of an OPE parameterization~\cite{Falk:1998jq} which follows the $P^*_{min}$ dependence of the measured moments.}
\end{figure}
\newpage

\begin{figure}[H]
\begin{center}
\begin{picture}(15.,13.4)
\put(0.0,0.0){\mbox{\epsfig{file=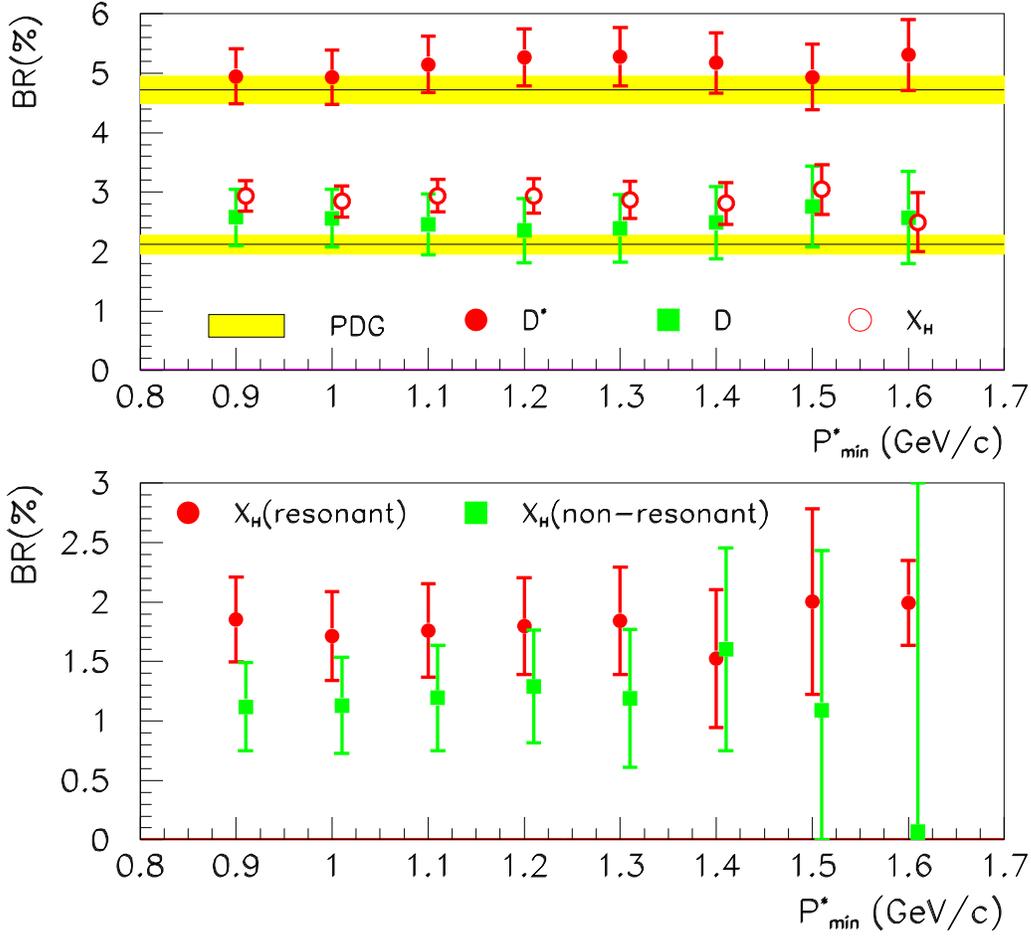, scale=1.0}}}
\end{picture}   
\end{center}
\vspace{-1.0cm}
\caption{\em \label{br_frac} The upper figure shows the measured branching fractions of $D^*$, $D$ and $X_H$ as a function of $P^*_{min}$ that are obtained from the $M_X$ fits after acceptance correction (points), and the world average~\cite{PDG} branching fractions for $D^*$ and $D$ (yellow bands). The lower figure shows the extracted branching fractions obtained from a fit where the resonant and non-resonant components are allowed to float independently. All branching fractions are obtained by assuming a total semileptonic branching fraction of 10.87\%.}
\end{figure}

The first moment of $M_X^2$ (or any other moment) can be extracted from the fitted relative contributions of the different decay modes, $f_{D^*}$, $f_D$, and $f_{X_H}$.  These fractions are normalized to the $B\ra D^* \ell \nu$ branching fraction, and as a result many uncertainties in the detection efficiencies and branching fractions cancel. 
Taking into account the true particle masses (the $D$ and $D^*$ masses are basically $\delta$ functions, and the mean of the $X_H$ contribution is taken from generated events) the first moments are calculated according to the following expression
\begin{equation}
\label{mxequation}
\langle M_X^2 - \overline{m}_D^2 \rangle = f_{D^*} \cdot (M_{D^*}^2-\overline{m}_D^2) + f_{D} \cdot (M_{D}^2-\overline{m}_D^2) + f_{X_H} \cdot \langle M_{X_H}^2-\overline{m}_D^2 \rangle .
\end{equation}
The results obtained for $\langle M_X^2 - \overline{m}_D^2 \rangle$ from the $\chi^2$ fit are summarized in Table \ref{tab:mom} and Figure~\ref{moments_ivar2} as a function of the minimum lepton momentum $P^*_{min}$, ranging from $0.9 - 1.6 \gevc$. 
The errors for different $P^*_{min}$ are highly correlated because all events selected for a certain $P^*_{min}$ are contained in samples with lower  $P^*_{min}$. The CLEO Collaboration has also measured the first hadronic moment for $P^*_{min} =1.5~\gevc$. Their result of  $\langle M_X^2 - \overline{m}_D^2 \rangle = 0.251 \pm 0.066~[\gevcc]^2$~\cite{Cronin-Hennessy:2001fk} is  
compatible with the measurement presented here.\\


\section{Studies of Systematic Uncertainties}

\label{sec:Systematics}
Extensive studies have been performed to assess the systematic uncertainties in the measurement of the first hadronic moment as a function of $P^*_{min}$. These studies involve both changes in the event selection and variations of the various corrections for efficiencies and resolution, and comparisons of data with Monte Carlo simulations for various control samples. The systematic errors are listed in Table \ref{tab:mom}.

With the current data set the total error on the moments is statistically dominated. One of the most important systematic errors comes from the uncertainty in the model for the higher mass states $X_H$. The resonant (ISGW2) and non-resonant (Goity-Roberts) decays are each modeled for four decay modes. The systematic error is evaluated by alternately turning on and off all individual components of the two models and choosing all possible combinations of decay modes for each of the two models. In this way not only the shape of the resonant and non-resonant component is varied but also the relative branching fractions between the different high-mass states and their true underlying masses are altered. We assign the r.m.s. deviation of these variations as the systematic error due to the $X_H$ modeling. This error has the largest contribution for low $P^*_{min}$ and decreases for higher lepton momenta because the high-mass final states are significantly reduced. 

Another systematic error arises from differences in the track and photon resolution between Monte Carlo simulations and control samples in data. The effect on the moments is estimated by varying the resolution and changing the bin width of the $M_X$ distribution.  

The third important systematic error is due to the uncertainties in the background contribution from secondary semileptonic decays of charm particles.
This background was varied by up to $\pm 30 \%$ in steps of $5\%$ and the $\chi^2$-probability weighted r.m.s. deviation of these variations is taken as the systematic error. This error also decreases for higher $P^*_{min}$. The $M_X$ distribution obtained from wrong-sign leptons in data and Monte Carlo simulation is found to be in good agreement. 
The systematic uncertainty due to the sideband subtraction is evaluated by varying the lower border of the signal window from $5.265 - 5.275~\gevcc$ and reducing the sideband region from the default size of $5.20 - 5.27~\gevcc$ to $5.24 - 5.27~\gevcc$. This addresses possible $M_{ES}$ dependent changes in the shape of the $M_X$ distribution obtained from the sideband as well as uncertainties regarding the ratio of background in the signal region to background in the sideband region. The full variation of the moments was taken as the systematic error.

The impact of changing the $M_X$ range for the fit was also studied and the variation of the measured moments is taken as a systematic error. Its size is almost $P^*_{min}$ independent and is included in the error quoted for detector systematics.

\begin{figure}[H]
\begin{center}
\begin{picture}(15.,20.4)
\put(0.0,0.0){\mbox{\epsfig{file=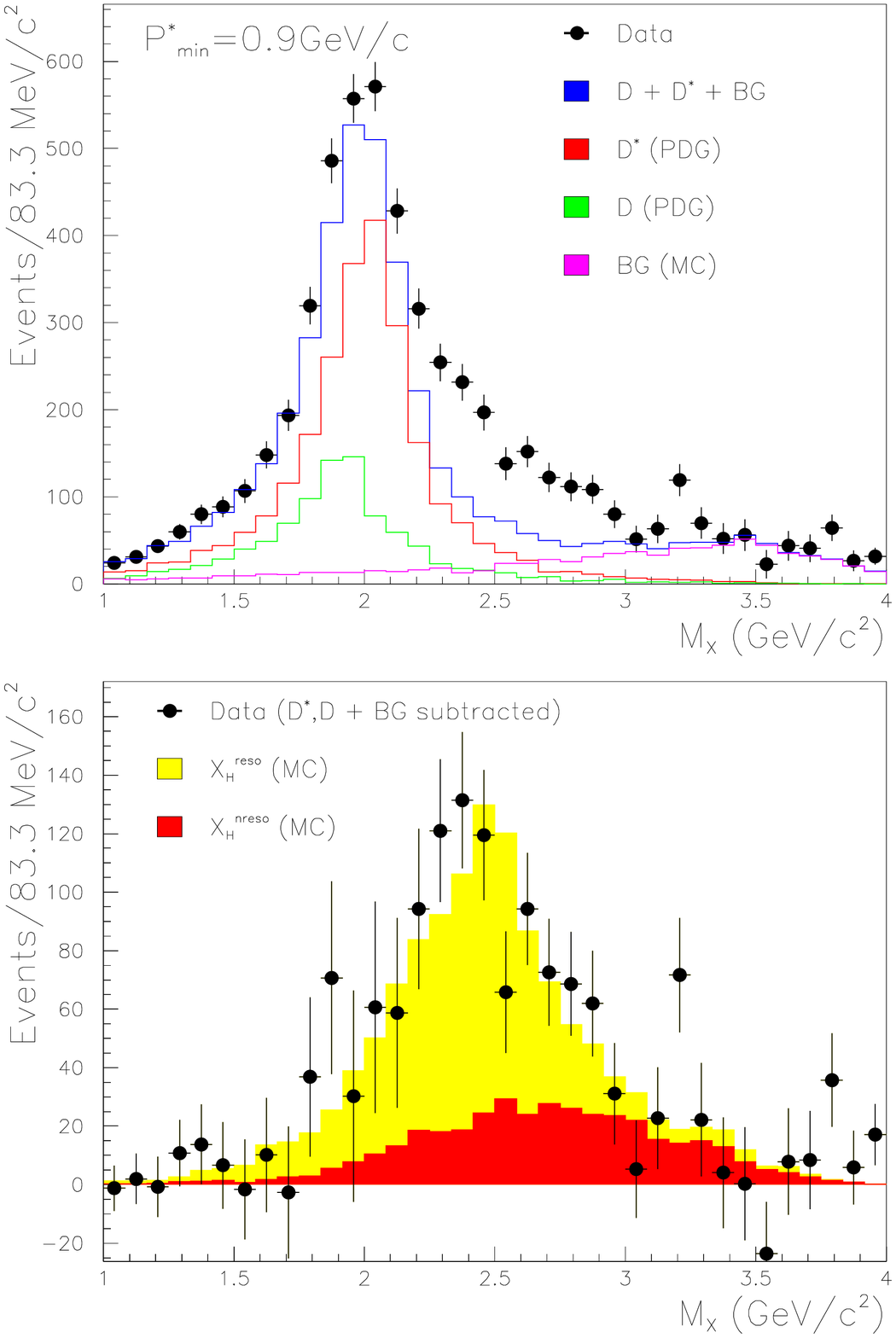, scale=0.75}}}
\end{picture}   
\end{center}
\vspace{-1.cm}
\caption{\em \label{xhdirect} The upper figure shows the sideband-subtracted $M_X$ distribution along with the expected $D$, $D^*$ and background components for $P^*_{min}=0.9~\gevc$. The lower figure shows the $M_X$ distribution of the high-mass final states obtained after subtracting the $D$, $D^*$, and background components. The filled histograms correspond to the Monte Carlo expectations for the resonant (grey) and non-resonant (black) components.}
\end{figure}

Several cross checks were carried out. In particular, we performed  a model independent, direct measurement of $M_X$ for the resonant and non-resonant states by subtracting the known $D$, $D^*$ contributions and background from the $M_X$ distribution of the data. The branching fractions for $D$ and $D^*$ were taken from the PDG~\cite{PDG} and the $P^*_{min}$ dependent acceptance of both decay modes as well as the background were taken from the Monte Carlo simulation. Figure~\ref{xhdirect} shows the $M_X$ distribution before and after subtraction for the lowest $P^*_{min}$ of $0.9~\gevc$. 
\begin{figure}[H]
\begin{center}
\begin{picture}(15.,11.4)
\put(1.0,0.0){\mbox{\epsfig{file=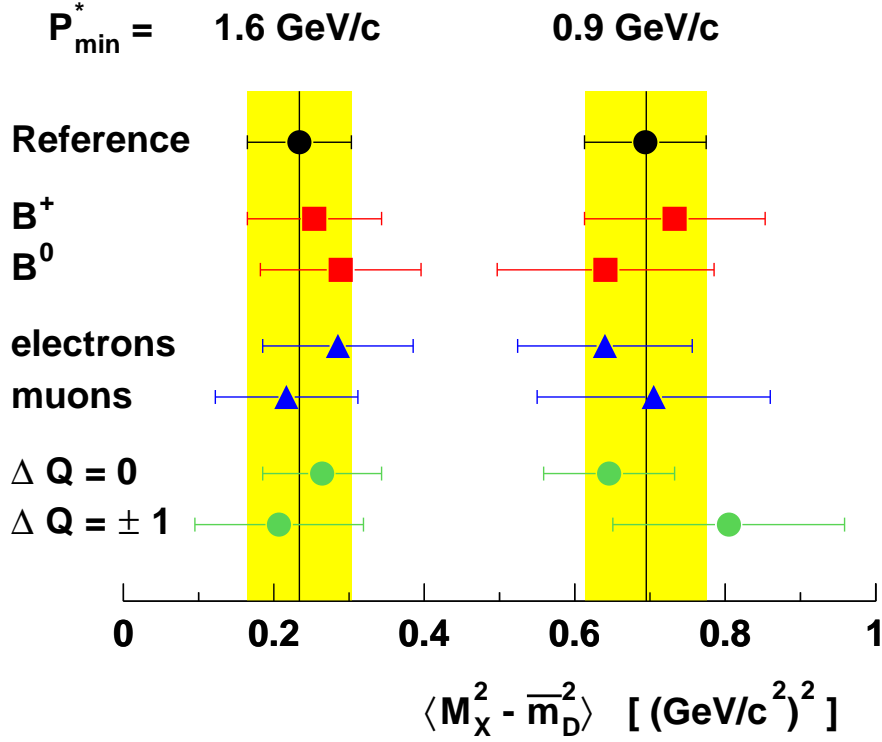, scale=0.7}}}
\end{picture}   
\end{center}
\vspace{-1.5cm}
\caption{\em \label{ursl} A comparison of the results for the first hadronic moment obtained from pairwise independent subsamples.}
\end{figure}
A significant contribution to the $M_X$ distribution from high-mass final states can be observed and is found to be in good agreement with the  Monte Carlo simulation. This measurement of the high-mass states is not corrected for detector resolution effects. However, through the use of the two-constraint kinematic fit, the mean is believed to be an almost unbiased estimator of its true underlying value (see Figure~\ref{scanMmiss_vcb}). The mean of the measured mass distribution is in agreement with the mean $\langle M_{X_H} \rangle$ of the Monte Carlo distribution used in the standard extraction of the moments (see Eq.~\ref{mxequation}).

We also used the relative contributions $f_i$ to determine branching fractions by correcting for acceptance for each $P^*_{min}$ value and assuming a total semileptonic branching fraction of 10.87\%~\cite{thorsten}.
The dependence of these branching fractions on $P^*_{min}$ is shown in Figure~\ref{br_frac}. We obtain branching fractions for $D$ and $D^*$ that are both compatible with previous measurements~\cite{PDG} and stable as a function of $P^*_{min}$, although of course correlated. 

In order to further explore the non-resonant contributions to the $M_X$ distribution we split the combined shape used for the $X_H$ system into its resonant and non-resonant components and repeat the fits at different $P^*_{min}$ with four independent contributions. The branching fractions obtained by this means are also shown in Figure~\ref{br_frac}.

With the current data sample the sensitivity to the four independent contributions is not sufficient at large $P^*_{min}$  where the non-resonant contribution is suppressed. We find stable results for the resonant and non-resonant branching fractions and the ratio between the two is compatible with that used in the Monte Carlo simulation.

Other tests include a comparison of the moment measurements for different subsamples of the data: $B^{\pm}$ versus $B^0$, decays with electrons versus muons, and events with total charge $Q_{tot}= 0$ and $|Q_{tot}| = 1$. A comparison of these results is shown in Figure~\ref{ursl}.

The effect of variation of the signal-to-background ratio in the $B_{reco}$ samples was also studied. Likewise, the requirement on $|M_{miss}^2|$ was varied.
All results were found to be in good agreement with the standard extraction of the moments for the full data sample.

\section{Conclusions}
\label{sec:Physics}

We have performed a preliminary measurement of the first moment $\langle M_X^2 - \overline{m}_D^2 \rangle$ of the hadronic mass distribution in semileptonic $B$ decays and extend the measurements to lepton momenta as low as $0.9 \gevc$. As a result, these measurements are sensitive to higher-mass charm states, in the form of both resonant and non-resonant contributions. The full reconstruction of one of the two $B$ mesons and the kinematic fit to the full event have resulted in measurements that are limited by statistical uncertainties.
For lepton momenta above 1.5 \gevc, the measured moment agrees well with an earlier result reported by the CLEO Collaboration~\cite{Cronin-Hennessy:2001fk}.

Using the OPE, the measurements of $\langle M_X^2 - \overline{m}_D^2 \rangle$ can be translated into values for $\lambda_1$ and $\bar{\Lambda}$. The coefficients used in the expansions are obtained from calculations by Falk and Luke~\cite{Falk:1998jq,flpc}.  
The observed dependence of the moments on the minimum lepton momentum
can be reproduced by the OPE, as long as we adjust the non-perturbative parameters (see Figure~\ref{moments_ivar2} for an example).

If, on the other hand, we insert into the OPE calculations recent independent measurements of $\bar{\Lambda}$, the data cannot be reproduced. 
The CLEO Collaboration reported a measurement of $\bar{\Lambda} = 0.35 \pm 0.08 \pm 0.10~\gev$ \cite{Chen:2001fj}, derived from the first moment of the photon energy spectrum above 2.0~\gev in $b \rightarrow s \gamma$ transitions. 
CLEO also recently reported a measurement of $\bar{\Lambda}$ from the lepton energy spectrum in semileptonic $B$ decays, for $P^*_{min}=1.5~\gevc$, which yielded a comparable value \cite{cleolept}.
For $P^*_{min}=1.5 \gevc$
and $\bar{\Lambda}= 0.35\pm 0.13 \gev$,
we find
\begin{equation}
   \lambda_1 = -0.17 \pm 0.06 \pm 0.07 ~\gev^2 ,
\end{equation}
where the first error is experimental and the second theoretical.
As stated above, this result is also in good agreement with the CLEO measurement of $\lambda_1 = -0.236 \pm 0.071 \pm 0.078~\gev^2$.
However, when we use the same value of $\bar{\Lambda}$ and the moments measured over a wider lepton momentum range, we find values for $\lambda_1$ that are not compatible with each other. Specifically, we obtain
\begin{equation}
   \lambda_1|_{_{P^*_{min} = 0.9 \gevc}} - \lambda_1|_{_{P^*_{min} = 1.5 \gevc}} =  0.05~\gev^2 - (-0.17~\gev^2) = 0.22 \pm 0.04 \pm 0.05~\gev^2.                       
\end{equation}
The first error is statistical and takes into account the correlation between the individual moment measurements. The second error is systematic and is dominated by the uncertainty on~$\bar{\Lambda}$.

Another way to look at this is to use $\lambda_1 = -0.17\gev^2$ and $\bar{\Lambda} = 0.35 \gev$ and calculate~\cite{Falk:1998jq,flpc} the moments as a function of $P^*_{min}$. We find a much smaller momentum dependence than the data indicate (see Figure~\ref{moments_ivar2}). 
This suggests that the observed  contribution from the high-mass states, i.e., the non-resonant $X_H^{nreso}$ decays, is very important.

In summary, for high lepton momenta, the moment measurements confirm earlier results by the CLEO Collaboration, and  using $\bar{\Lambda} = 0.35 \pm 0.13~\gev$ we obtain $\lambda_1 = - 0.17~\pm 0.06 \pm 0.07 \gev^2$ for $P^*_{min} = 1.5 \gevc$.
However, based on the OPE predictions for the $P^*_{min}$ dependence of the hadronic moments these values of the parameters fail to describe the moment measurements variation to lower lepton momenta.
This inconsistency suggests that some of the underlying assumptions based on HQET and the OPE require further scrutiny.

\section{Acknowledgments}

\label{sec:Acknowledgments}


The authors wish to express their appreciation to A. Falk and M. Luke for providing the coefficients for extracting $\lambda_1$ and $\bar{\Lambda}$. They would like to thank them as well as Z. Ligeti and Ch. Bauer for valuable discussions regarding the interpretation of the results.


We are grateful for the 
extraordinary contributions of our \pep2\ colleagues in
achieving the excellent luminosity and machine conditions
that have made this work possible.
The success of this project also relies critically on the 
expertise and dedication of the computing organizations that 
support \babar.
The collaborating institutions wish to thank 
SLAC for its support and the kind hospitality extended to them. 
This work is supported by the
US Department of Energy
and National Science Foundation, the
Natural Sciences and Engineering Research Council (Canada),
Institute of High Energy Physics (China), the
Commissariat \`a l'Energie Atomique and
Institut National de Physique Nucl\'eaire et de Physique des Particules
(France), the
Bundesministerium f\"ur Bildung und Forschung and
Deutsche Forschungsgemeinschaft
(Germany), the
Istituto Nazionale di Fisica Nucleare (Italy),
the Research Council of Norway, the
Ministry of Science and Technology of the Russian Federation, and the
Particle Physics and Astronomy Research Council (United Kingdom). 
Individuals have received support from 
the A. P. Sloan Foundation, 
the Research Corporation,
and the Alexander von Humboldt Foundation.

\bibliography{refs}         

\bibliographystyle{h-physrev2-original}   %

\end{document}